\documentclass[a4paper]{spie}  
\usepackage[]{graphicx}

\title{New results on focusing of gamma-rays with Laue lenses} 

\author{F. Ferrari\supit{a}, F. Frontera\supit{a,c}, G. Loffredo\supit{a}, E. Virgilli\supit{a}, 
C. Guidorzi\supit{a}, V. Carassiti\supit{b}, F. Evangelisti\supit{b}, L. Landi\supit{a}, 
S. Chiozzi\supit{b},  S. Squerzanti\supit{b}, E. Caroli\supit{c}, J.B. Stephen\supit{c}, F. Schiavone\supit{c},
A. Basili\supit{c}, K.H. Andersen\supit{d}, P. Courtois\supit{d}
\skiplinehalf
\supit{a}University of Ferrara, Physics Department, Via Saragat 1, 44100
Ferrara, Italy; \\
\supit{b}Istituto Nazionale Fisica Nucleare, Sezione di Ferrara, Via Saragat 1,
44100 Ferrara, Italy; \\
\supit{c}INAF, IASF Bologna, Via Gobetti 101, 40129 Bologna, Italy\\
\supit{d}Institute Laue--Langevin, 6 Rue Jules Horowitz, 38042 Grenoble, France\\
}

\authorinfo{Further author information: (Send correspondence 
to F.F.)\\ F.F: E-mail: frontera@fe.infn.it, Telephone: +39 0532 974 254}

\begin{document} 
  \maketitle 

\begin{abstract}
We report on new results on the development activity of broad band Laue 
lenses for hard X-/gamma-ray astronomy (70/100-600 keV). After the development
of a first prototype, whose performance was presented at the SPIE conference 
on Astronomical Telescopes held last year in Marseille (Frontera et al. 2008), 
we have improved the lens assembling technology. We present the 
the development status of the new lens prototype that is on the way to be assembled.

\end{abstract}


\section{INTRODUCTION}
\label{s:intro}  

Along with the hard X-ray astronomy up to 70/100 keV, also the gamma--ray astronomy 
above this energy is moving from direct sky-viewing telescopes to focusing telescopes.
With the advent of focusing telescopes in this energy range, a 
big leap is expected, in both sensitivity and angular resolution.
As far as the sensitivity is concerned, the expected increase is by a 
factor of 100-1000 with respect to the best non-focusing instruments of 
the current generation
(e.g., BeppoSAX/PDS, Ref.~\citenum{Frontera97}; INTEGRAL/IBIS, Ref.~\citenum{Ubertini03}).
Concerning the angular resolution, the increase is expected by more than a factor 10 
(from $\sim 10$ arcmin of the mask telescopes like INTEGRAL IBIS to 
less than 1 arcmin).
The next generation of  gamma--ray ($>$70/100 keV) focusing telescopes will make use 
of the Bragg diffraction technique from mosaic-like crystals in transmission configuration 
(Laue lenses). 
The expected astrophysical issues that are expected to be solved with the advent
of these telescopes are many and of fundamental importance. A discussion of them is 
done in the context of the mission proposal {\em Gamma Ray Imager} (GRI), submitted 
to ESA in response to the first AO of the 'Cosmic Vision 2015--2025' plan (Ref.~\citenum{Knodlseder07}). 
For the astrophysical importance of the $>$100 keV band, see also
Refs.~\citenum{Frontera05a,Frontera06,Knodlseder06,Knodlseder07}. 

Here we report on the status of 
our project HAXTEL (= HArd X-ray TELescope) devoted to  developing the technology 
for building Laue lenses with broad energy passband (70/100--600 keV).
The results of this development activity over the last few years
have been reported and discussed 
(see Refs.~\citenum{Frontera06,Frontera07} and references therein) with a summary of them 
given in Ref.~\citenum{Frontera08}.

Last year we reported the first lens Prototype Model (PM) and its performance
(see Ref.\citenum{Frontera08}). 
Before discussing the activity performed this year, we summarize the main features
of the assembly technique and the  performance of the first PM.

\section{The lens assembly technique}

The technique adopted  is based on the use of a counter-mask 
provided with holes, two for each crystal tile. Each tile is positioned
on the countermask by means of two cylindrical pins, rigidly glued to the tile, that
are inserted in the countermask holes. The pin direction and the axis
of the average lattice plane of each crystal tile have to be exactly
orthogonal. The hole axis direction constrains the energy
of the photons diffracted by the tile, while the relative position of
the two holes in the countermask
establishes the azimuthal orientation of the mean crystal lattice plane,
that has to be othogonal to the lens axis.
Depending on the countermask shape and, mainly, on the direction of the hole 
axes, the desired geometry of lens can be obtained. 
In the case of a lens for space astronomy (i.e., parallel fluxes), the hole axes 
have to be all directed toward the center of curvature of the lens. 

In the case of the developed PM, the hole axes are set parallel to the
of the lens axis. This choice has been made to illuminate the entire lens with the
available test gamma--ray beam, which is isotropical, being generated by an X--ray
tube, and also highly divergent, as the source is only  a small distance 
away ($d \sim 6$~m) from the lens.

The adopted mosaic crystals are made of Cu with chosen lattice planes (111).
Once the average direction of the chosen lattice planes of
each crystal tile has been determined, the two pins, inserted in a pin
holder, are glued to the crystal. The pin holder direction is first
made parallel to the gamma--ray beam axis and thus to the direction of the 
average crystalline plane chosen (in our case, the (111) planes). 

Once all the crystal tiles are placed on the counter-mask, a
frame is glued to the entire set of the crystals. Then the lens frame, along
with the crystals, is separated 
from the counter-mask and from the pins by means of a chemical etching that 
dissolves the aluminum caps that cover the pin bases glued to each crystal.

\begin{figure}
\begin{center}
\includegraphics[angle=0, width=0.4\textwidth]{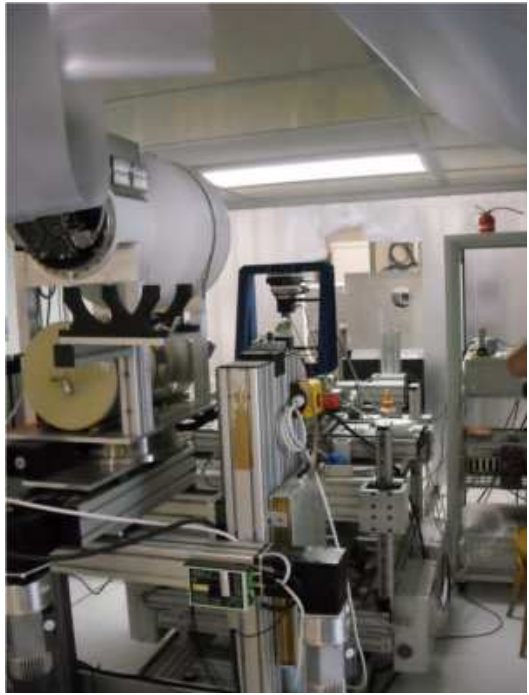}
\end{center}
\caption{A view of the apparatus  located in the new clean room for the lens 
assembling. The apparatus is installed in the LARIX laboratory of the University
of Ferrara.}
\label{f:facility}
\end{figure}

The average direction of the chosen crystalline planes and of the pin axes
are determined by means of an X--ray beam developed for this project and located
in the LARIX (LArge Italian X--ray facility) laboratory of the University of Ferrara.
For a LARIX description see Ref.~\citenum{Loffredo05}. 
A view of the experimental apparatus in the current configuration 
is shown in Fig.~\ref{f:facility}. 
%
%
As discussed in Ref.~\citenum{Frontera07}, the developed apparatus includes 
an X--ray generator tube with a fine focus of 0.4 mm radius, a  maximum 
voltage of 150~kV and a maximum power of $\sim 200$~W.

The photons coming out from the gamma--ray source box
are first collimated, with the beam axis made horizontal and directed toward the
centre of a large (30 cm diameter) X--ray imaging detector with image pixels 
of 300~$\mu$m.
The collimator aperture can be remotely adjusted in two orthogonal directions
in order to have the desired X--ray beam size (see Ref.~\citenum{Frontera07}).   
In addition to the X--ray imaging detector, a cooled HPGe detector and 
a position sensitive (2 mm) scintillator detector are available.

For each mosaic crystal tile, the direction of the average lattice plane 
can be  determined with an accuracy better than 10 arcsec (see Ref.~\citenum{Frontera07}).

The first developed PM was composed of a ring of 20 mosaic crystal tiles of Cu(111)
with a ring diameter of 36 cm. The 
mosaic spread  of the used crystals ranged from $\sim 2.5$ and $\sim 3.5$ arcmin.
The tile cross--section was 15$\times$15 mm$^2$ while its thickness was 2 mm. 
The lens frame was made of carbon fiber of 1 mm thickness.

The PM was widely tested using the polychromatic X--ray beam described above, as reported
in the Ref.~\citenum{Frontera08}.

\begin{figure}
\begin{center}
\includegraphics[angle=0, width=0.4\textwidth]{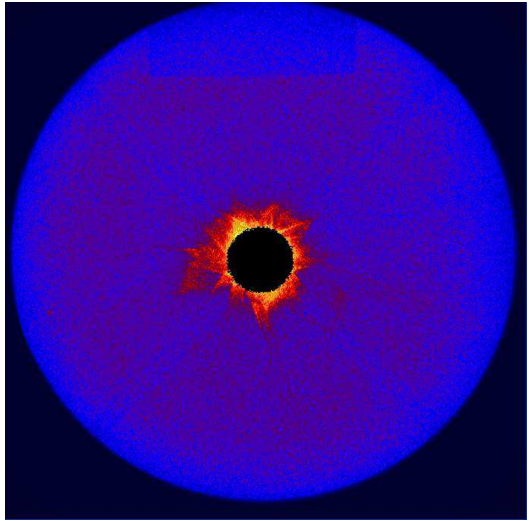}
\end{center}
\caption{Difference between the measured PSF 
and that obtained with a Monte Carlo code by assuming a perfect positioning 
of the crystal tiles in the lens.}
\label{f:difference_image}
\end{figure}

The difference between the measured PSF 
and the simulated one is shown in Fig.~\ref{f:difference_image}.
As can be seen, only the center part of the measured image (the black region) 
is  subtracted by the simulated image. The corona still visible
in the difference image is mainly the result of the cumulative error made in the 
crystal tile positioning.
%
%

The disagreement between the measured and the expected PSF is clear.
It can be seen (see Ref.~\citenum{Frontera08} for details) that at the radius at which the expected
distribution of the focused photons reaches the saturation, only  60\% of the focused 
photons are collected.

\section{Error budget analysis and new improvements}

By shielding all the lens cystals but one, we have investigated 
the contribution of each lens crystal to
the lens PSF and to the reflected photon spectrum. From these results, we have 
derived the positioning error of each crystal in the lens.
In order to establish the true contribution of each assembling step to the 
cumulative error, we have performed further 
tests. In this way we have estimated the precision achieved in each 
assembling step.

On the basis of this analysis, we have started a new design of the lens 
assembling apparatus.
The major changes have concerned the following parts.

\subsection{New pin and crystal holders}

\begin{figure}
\begin{center}
\includegraphics[angle=0, width=0.4\textwidth]{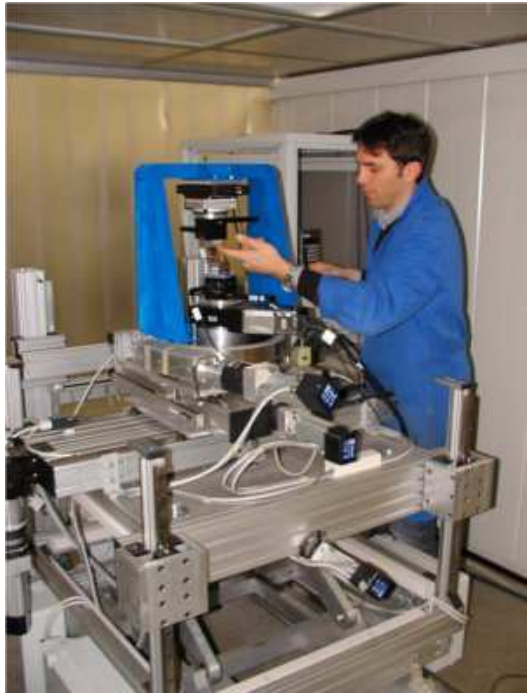}
\end{center}
\caption{New pin holder while it is mounted in the lens assembling apparatus.}
\label{f:new-portaspine}
\end{figure}

A new pin holder has been developed (see Fig.~\ref{f:new-portaspine}). As the previous
one (see Ref.~\citenum{Frontera07}), it can be rotated along a circle with its center in the  
vertical axis of the crystal holder, and can be tilted along two 
orthogonal directions as the crystal holder. However, while the alignment of the old pin
holder to the beam axis was obtained by using the
X--ray shadow projected by two Tungsten crosses located along the pin positioner,
now we use a Silicon crystal in transmission configuration with the chosen lattice
planes (220) that are othogonal to the crystal surface within a few arcsec. 
The pins are located in mechanically worked holes with their axis orthogonal to the
crystal surface. This angle is known with a precision better than 1 arcmin.
Using the Bragg diffraction we can determine the direction of the chosen Si 
lattice planes (within a few arcsec) and thus perform the alignment of the pins to the
gamma-ray beam axis.

\begin{figure}
\begin{center}
\includegraphics[angle=0, width=0.4\textwidth]{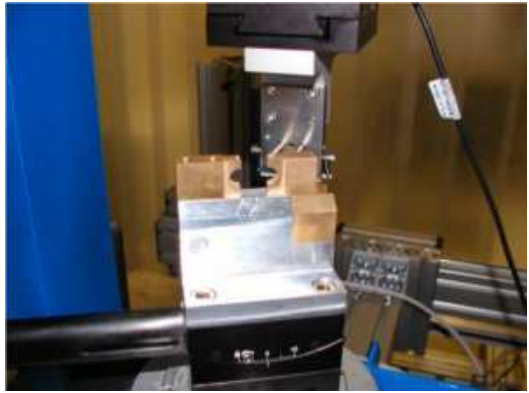}
\end{center}
\caption{New crystal holder.}
\label{f:new-crystalholder}
\end{figure}

Also the crystal holder (see Fig.~\ref{f:new-crystalholder})
has been improved with respect to the previous one (Ref.~\citenum{Frontera07}), 
and now it can host crystals of different thickness.

%
%

%
%

\subsection{New pin geometry}

For the new lens prototype, half of the crystal tiles are positioned in the countermask
using a pair of pins with cylindrical shape, as in the first prototype
(see Ref.~\citenum{Frontera08}), while the other half of crystal 
tiles are positioned using new pins of conical shape with elliptical section. In this
case it is sufficient one pin for each tile.

\subsection{Clean room and temperature control of the lens assembling process}

For the realization of the new prototype, all the measurements and the assembling process are 
now performed in a clean room of about 50000 class and an active temperature control within 1~$^o$C. 
This temperature control is expected, from mechanical simulations, guarantee negligible 
distortions of the lens.

\subsection{New counter mask and lens support}

\begin{figure}
\begin{center}
\includegraphics[angle=0, width=0.4\textwidth]{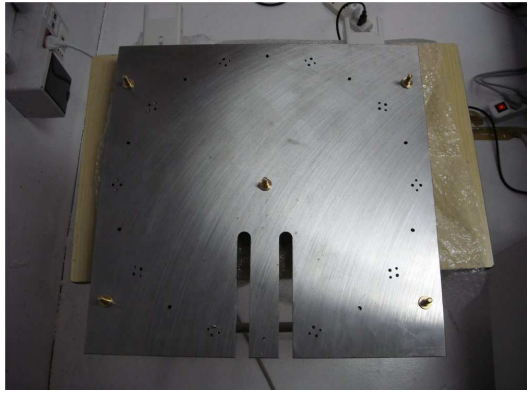}
\end{center}
\caption{New countermask.}
\label{f:new-countermask}
\end{figure}

The new counter mask (see Fig.~\ref{f:new-countermask}) can allocate either cylindrical pins 
and conical pins with 
elliptical shape. 

\begin{figure}
\begin{center}
\includegraphics[angle=0, width=0.4\textwidth]{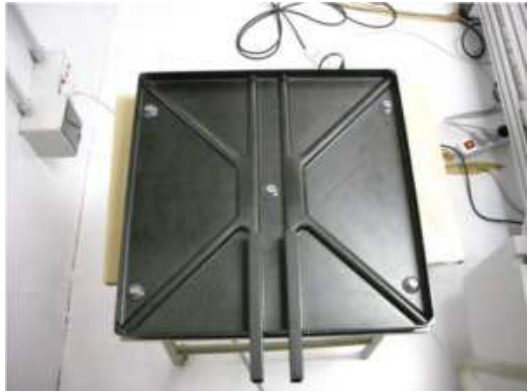}
\end{center}
\caption{New lens frame.}
\label{f:frame}
\end{figure}

The lens frame (see Fig.~\ref{f:frame}) is made of  8 layers of carbon fibers properly oriented
in order to optimize and get isotropical its thermal conductivity and stress,
thus avoiding distortions of the lens for temperature changes, that, in any case, should be held
within $\pm 2$~$^o$C (see above). 

%
%

%
%

\section{Alignment measurements}

All the alignments have been performed using the gamma-ray beam 
available in the LARIX facility. For its properties see Frontera et al. 2008 
(Ref~\citenum{Frontera07}).

\subsection{Alignment of the pin holder to the gamma-ray beam}

\begin{figure}
\begin{center}
\includegraphics[angle=0, width=0.4\textwidth]{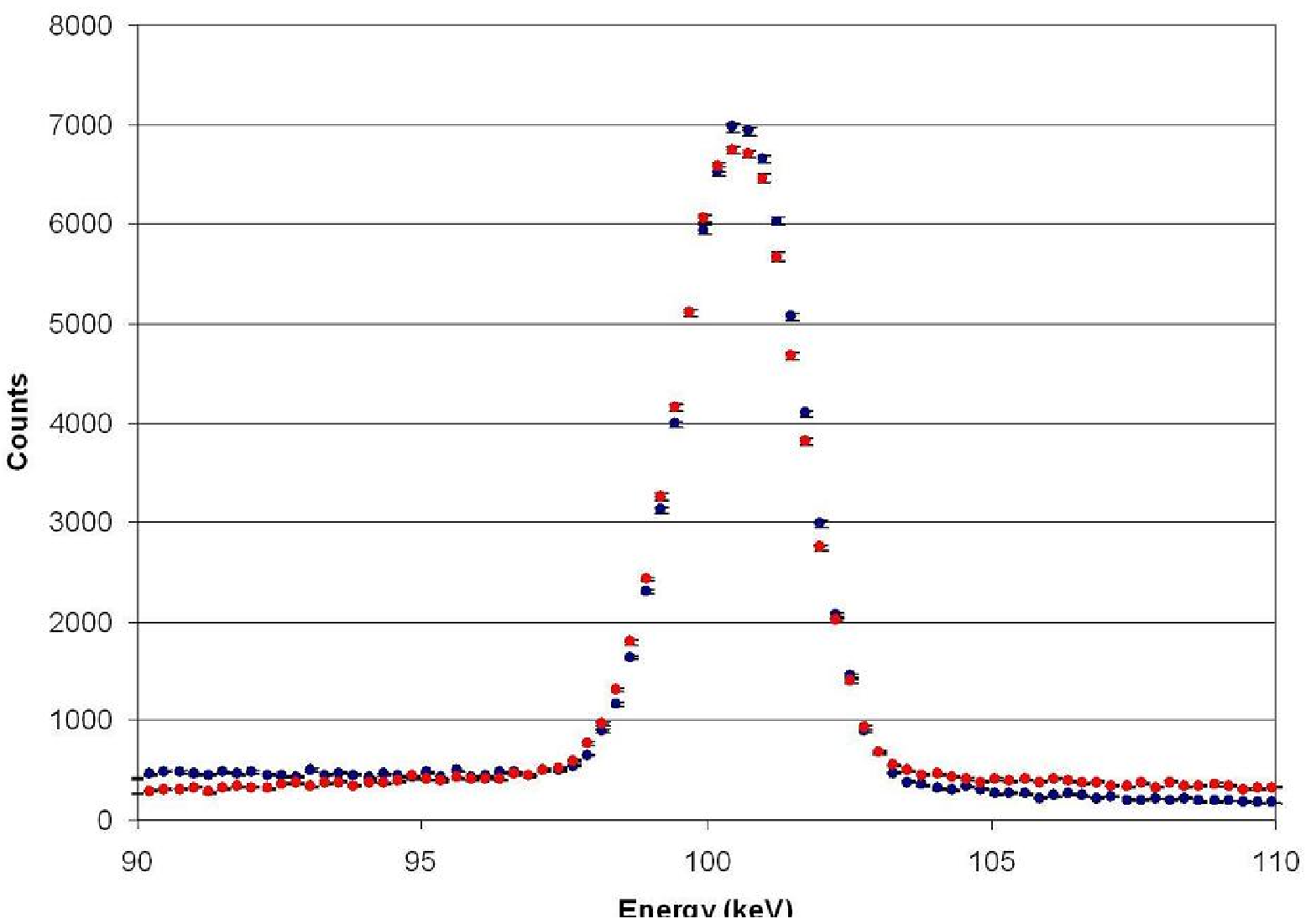}
\includegraphics[angle=0, width=0.4\textwidth]{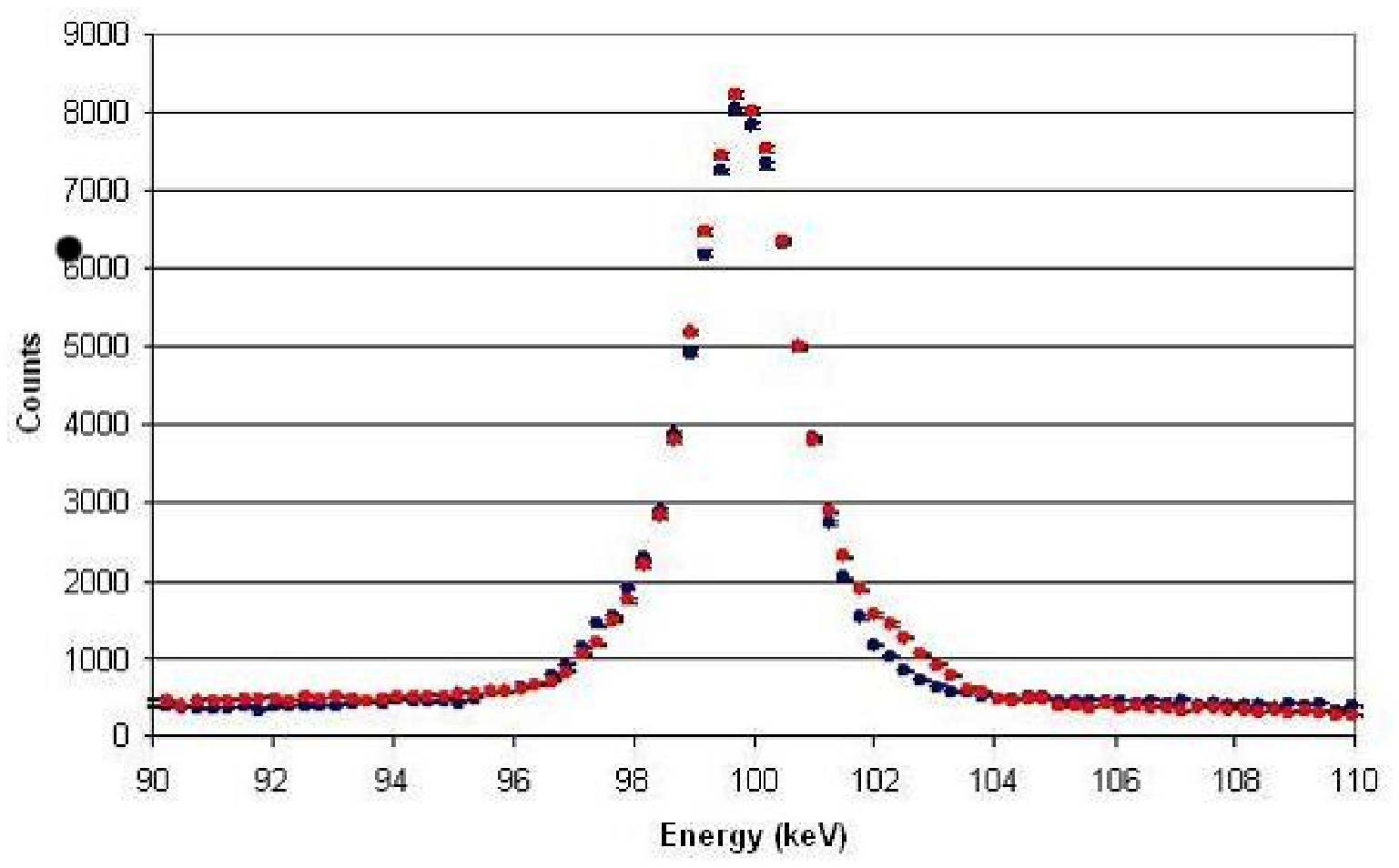}
\end{center}
\vspace{-0.5cm}
\caption{{\em Left panel}: Silicon diffracted spectra at equal left and right angles with respect to
the beam axis. {\em Right panel}: Silicon diffracted spectra at equal angles upward and downward 
with respect to the beam axis.}
\label{f:si_le-ri-up-down}
\end{figure}

\begin{figure}
\begin{center}
\includegraphics[angle=0, width=0.4\textwidth]{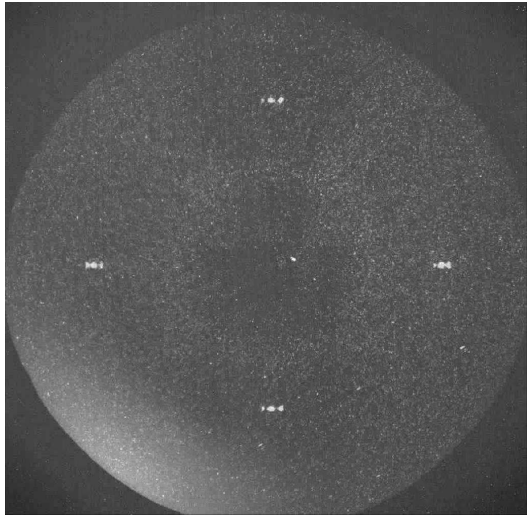}
\end{center}
\caption{Image of the Si (220) perfect crystal for diffraction of the gamma--ray
beam at symmetrical angles (left and right; upward and downward)  
with respect to the beam axis. The imager space resolution is 0.3 mm}
\label{f:si_thales}
\end{figure}

Taking into account the above description, the pin axes were made parallel to the
beam exploiting the Si crystal, imposing that the diffracted beam at equal left and right 
angles, and at equal angles upward and downword,  with respect to 
the beam direction,  give coincident spectra. The final results for the
pin alignments are shown in Figs.~\ref{f:si_le-ri-up-down}. 
When observed with an X--ray imager, the corresponding diffracted images 
are shown in Fig.~\ref{f:si_thales}.

%
%
%
%

%
%

\subsection{Lens crystals, their average lattice plane direction determination and pin pasting}

Once the axis of the pin holder  has been made parallel to the beam axis, the lattice plane
of each crystal has been determined and either a couple of cylindrical pins or one conical 
pin is glued to the crystal. This activity is being carried out.

\begin{figure}
\begin{center}
\includegraphics[angle=270, width=0.4\textwidth]{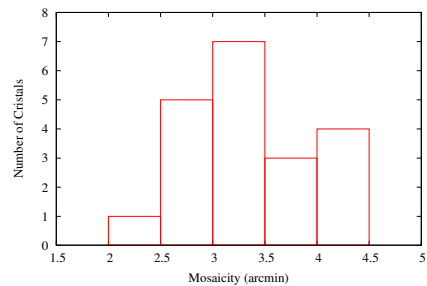}
\end{center}
\vspace{-0.5cm}
\caption{Distribution of the mosaic spread of the Cu(111) crystals that will be
used for the new Laue lens prototype.}
\label{f:spread}
\end{figure}

As for the previous prototype the crystals used for this new lens prototype are made
of Cu (111) with mosaic structure. The thickness all the crystal tiles is 3 mm and their
cross section is 15$\times$15~mm. The distribution of the mosaic spread of the 20 selected
crystals is shown in Fig.~\ref{f:spread}. 

\begin{figure}
\begin{center}
\includegraphics[angle=0, width=0.4\textwidth]{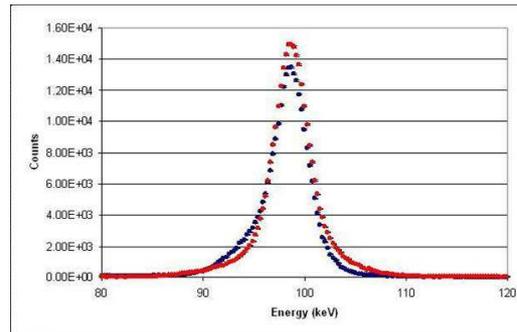}
\end{center}
\vspace{-0.5cm}
\caption{Example of spectral response of a Cu(111) mosaic crystal tile when the X--ray
beam is diffracted by the crystal at symmetrical angles with respect to the beam axis direction.}
\label{f:sampleCu}
\end{figure}

An example of spectral response of a crystal tile, once its average lattice plane is determined
and made parallel to the pin axis is shown in Fig.~\ref{f:sampleCu}, after two cylindrical pins 
have been pasted to it.

%
%
%
%
%
%

\section{Conclusions}

After the development and testing of the first Laue lens prototype (see Ref.~\citenum{Frontera08}),
we have devoted a new effort to improve the lens focusing quality.
In this paper we have described the new upgrading of the lens assembling system, that is being
used for building a new lens prototype. This is expected to be ready in one--two months.
The test results will be reported soon.

\acknowledgments     
 
We acknowledge the financial support by the Italian Space Agency ASI. 
The design study was also possible 
thanks to the award of the 2002 Descartes Prize of the European Committee.



\bibliography{lens_biblio}   
\bibliographystyle{spiebib}   

\end{document}